\documentclass[11pt,twoside]{article}
\usepackage{newpasp}
\usepackage{epsf}

\index{Shull, J. M.}

\begin{document}

\title{Hubble and FUSE Studies of Ly$\alpha$ Absorbers at Low $z$}
\author{J. Michael Shull}
\affil{CASA and JILA, Department of Astrophysical
\& Planetary Sciences, University of Colorado and NIST,
Boulder, CO 80309-0389}

\begin{abstract}
Ultraviolet spectrographs aboard the {\it Hubble Space Telescope} (HST) 
and the {\it Far Ultraviolet Spectroscopic Explorer} (FUSE) have proved 
their value as sensitive probes of the low-density intergalactic
medium (IGM) at low redshifts ($z < 0.1$).  Recent observations 
in Ly$\alpha$, Ly$\beta$, and occasional higher Lyman lines show that 
warm photoionized gas in the low-$z$ IGM may contain 20--25\% of the 
baryons, with a N$_{\rm HI}^{-1.8}$ distribution in column density. 
Measurements of resonance lines of Si~III, C~III, C~IV, and O~VI 
suggest that the metallicity of these absorbers ranges 
from 1--10\% of solar abundance down to values below 
$0.003 Z_{\odot}$.  A comparison of Ly$\beta$/Ly$\alpha$
ratios (FUSE and HST) yields a distribution of Doppler parameters
with $\langle b \rangle = 31.4 \pm 7.4$ km~s$^{-1}$
and median 28 km~s$^{-1}$, comparable to values at $z$ = 2--3.
The curve-of-growth (CoG) $b$-values are considerably less than widths 
derived from Ly$\alpha$ profile fitting,
$\langle b_{\rm CoG} / b_{\rm width} \rangle = 0.52$, which suggests that
low-$z$ absorbers contain sizable non-thermal motions or
velocity components arising from cosmological expansion and infall.     
A challenge for future UV spectroscopic missions (HST/COS and SUVO) is 
to obtain precision measurements of $\Omega_{\rm IGM}$ 
and metallicities for the strong Ly$\alpha$ absorbers that dominate the 
IGM baryon content.  This program will require accurate determinations of: 
(1) curves of growth using higher Lyman series lines; 
(2) the ionizing radiation field at 1--5 Ryd; and 
(3) characteristic sizes and shapes of the absorbers.
\end{abstract}


\section{Introduction}

For sheer sensitivity to interstellar or intergalactic H~I, no
technique compares with absorption lines of Ly$\alpha$ (1215.670 \AA)
and Ly$\beta$ (1025.722 \AA).  Owing to its large dipole oscillator
strength, Ly$\alpha$ is sensitive to gas with column
density N$_{\rm HI} \approx 10^{12.3}$ cm$^{-2}$, a million times
lower than typically detected in 21-cm emission.  It is little
wonder that astronomers and cosmologists are drawn to Ly$\alpha$
studies as probes of the IGM, to investigate processes 
of galaxy formation, large-scale structure, IGM thermal history, and 
chemical evolution.   

To understand these processes, which characterized the epoch when the 
first stars, galaxies, and heavy elements were formed, requires large 
ground-based telescopes and high-resolution spectrographs.  
At redshifts $z > 1.5$, sufficient to redshift H~I Ly$\alpha$ 
into the visible, astronomers have thoroughly studied the 
``Ly$\alpha$ Forest" with the Keck/HIRES (Hu et al.\ 1995),
VLT/UVES (Kim, Cristiani, \& D'Odorico 
2001), and other spectrographs.  At lower redshifts, $z < 1.5$, one must
use ultraviolet telescopes to measure Ly$\alpha$. To make the
important connections between Ly$\alpha$ absorbers and signatures
of galaxy formation and large-scale structure, it is best to probe
the ``local Ly$\alpha$ forest" at $z < 0.1$, where galaxy surveys 
and optical/21-cm imaging can detect galaxies well below the nominal
$L^*$ limit.  It is this area of low-$z$ IGM studies that has interested
the Colorado group (Stocke et al.\ 1995; Shull et al.\ 1999a).

In its first several years, HST was used with the Faint Object
Spectrograph (FOS) to carry out the QSO Absorption Line Key Project 
(Bahcall et al.\ 1991, 1993; Jannuzi et al.\ 1998; Weymann et al.\ 1998).  
Among its primary results was a characterization of the 
Ly$\alpha$ forest at $z < 1.5$ at low resolution (230 km~s$^{-1}$) 
using strong Ly$\alpha$ lines (primarily with rest-frame equivalent 
width $W_{\lambda} > 240$ m\AA).  In this review, I discuss our 
moderate-resolution (19 km~s$^{-1}$) HST studies of the more numerous, 
weak Ly$\alpha$ lines ($W_{\lambda} \geq 10$ m\AA) using the Goddard 
High-Resolution Spectrograph (GHRS) and the Space Telescope Imaging 
Spectrograph (STIS). I also describe studies with the FUSE spectrograph, 
which has become a powerful probe of the IGM.  With similar resolution 
(15--25 km~s$^{-1}$) at wavelengths shortward of those accessible to 
HST and with LiF and SiC optics (Moos et al.\ 2000), FUSE provides 
access to the  spectroscopically rich far-UV band (912--1187 \AA) which 
contains Ly$\beta$, higher Lyman-series lines, and 
lines of key heavy elements, C~III (977.03 \AA) and O~VI
(1031.93, 1037.62 \AA).  

\begin{figure}[h]
\begin{center}
\epsfxsize=5.0in
\leavevmode
\epsffile{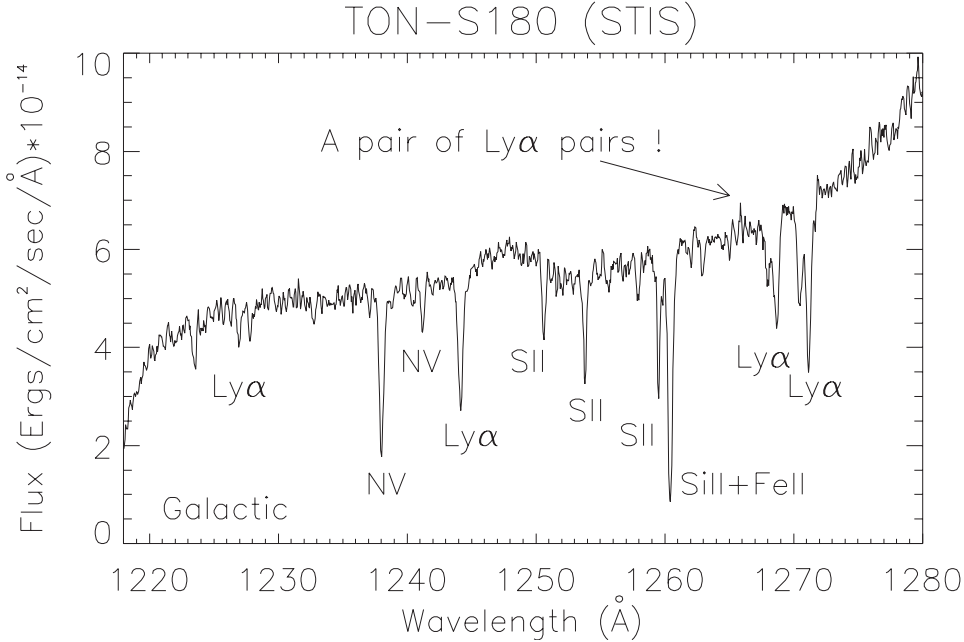}
\end{center}
\caption{\small
Typical STIS/G140M spectrum (Ton~S180) from Penton et al.\ (2001b)
showing a number of intergalactic
Ly$\alpha$ absorbers, including a ``pair of Ly$\alpha$ absorption pairs"
between 1267--1272 \AA.  Galactic interstellar lines of N~V, S~II, and
Si~II/Fe~II blend are also labeled.  }
\end{figure}

Figure 1 shows a typical moderate-resolution spectrum taken by STIS.
The Colorado survey results appear in a series of papers (Penton, Stocke,
\& Shull 2000; Penton, Shull, \& Stocke 2000; Penton, Stocke, \& Shull
2001a,b)   that describe an analysis of 197 absorbers along sightlines
to 31 AGN, over a total redshift pathlength $\Delta z = 1.15$.  Preliminary
FUSE results on the ``Ly$\beta$ forest" (Shull et al.\ 2000) have
been followed by a number of papers on individual IGM sightlines, including
PG~0953+415 (Savage et al.\ 2001), H1821+643 (Tripp et al.\ 2001),
and 3C~273 (Sembach et al.\ 2001).

The enormous amount of data on high-$z$ Ly$\alpha$ absorbers 
has constrained models for the growth of large-scale structure in the IGM.  
Similar comparisons can be made at low redshift, although the number of 
sightlines available for Ly$\alpha$ studies is far more limited.
Among the desired H~I statistics are the distribution in 
absorbers per unit redshift, $d{\cal N}/dz$, and their distribution in column 
density $N_{\rm HI}$.  We also wish to understand the degree of clustering, 
the two-point correlation function in velocity, and any associations of
absorbers with galaxies.  However, the redshift pathlength currently available 
for moderate-resolution studies ($\Delta z \approx 1$) is much 
smaller than that surveyed at high-$z$. This disparity should change with 
the installation of the {\it Cosmic Origins Spectrograph} (COS) on HST
in early 2004.  Another QSO Absorption-Line Key Project with COS
would be highly desirable.

\section{Low-Redshift IGM Studies}

\subsection{Scientific Goals}

It now appears likely that a substantial fraction of low-$z$ baryons 
reside in the IGM, distributed in comparable amounts between hot shocked 
gas (Cen \& Ostriker 1999; Dav\'e et al. 2001) and warm photoionized 
Ly$\alpha$ absorbers (Shull et al.\ 1996, 1999a).
At high redshifts, substantial progress has been made in linking these
absorbers with the lowest density fluctuations of the
baryon density; the statistics and physical properties of the absorbers
are interpreted in terms of global models for the gravitational collapse of
structure (Miralda-Escud\'e et al.\ 1996; Hernquist et al.\ 1996;
Zhang et al.\ 1997).  In a hierarchical dark matter-dominated
cosmological framework, the evolution of the Ly$\alpha$ forest
depends on the evolution of the dark matter and the thermal history of
the IGM (Croft et al.\ 1999; Hui \& Gnedin 1997; Schaye et al. 1999;
Ricotti, Gnedin, \& Shull 2000).   Dav\'e et al.\ (1999) show that 
the IGM overdensity--temperature relation (``effective equation of state'')
may be extended to low redshift and measured by the quantities $N_{\rm HI}$
and $b$.  We and others have pursued this idea theoretically
and have used HST data to test it (Ricotti et al.\  2000;
Dav\'e \& Tripp 2001).  However, until more observations are taken with
the HST/STIS echelle (7 km~s$^{-1}$ resolution), it will be difficult
to define the true thermal line widths and perform reliable thermodynamic
studies of the temperature evolution of the IGM.

With both FUSE and HST, our long-term goals are to characterize
the amount and distribution of baryons in the low-$z$ IGM and to 
define the extent of heavy-element transport in the IGM. 
Although many groups are continuing their studies of key individual 
sightlines, some of the most important long-term work involves
large-scale surveys of Ly$\alpha$, Ly$\beta$, and heavy elements. 
These surveys are intended to ``weigh the Ly$\alpha$ forest" (measure
$\Omega_{\rm IGM}$) and determine the history of IGM metal production 
and transport away from their sources.  

To perform these measurements accurately, one must detect both 
Ly$\alpha$ and Ly$\beta$,  as well as the accessible UV 
resonance lines such as Si~III $\lambda 1206.50$, Si~IV $\lambda
\lambda 1393.76, 1402.77$, C~III $\lambda 977.03$, 
C~IV $\lambda \lambda 1548.20, 1550.77$, and O~VI $\lambda \lambda 
1031.93, 1037.62$. Our work (Shull et al.\ 1999; Penton et al.\ 2000a,b) 
suggests that low-$z$ Ly$\alpha$ absorbers are an important gaseous
reservoir, with perhaps 25\% of the baryons remaining in the IGM 
from the epoch of galaxy formation.  However, because of the significant
photoionization corrections required for this estimate, large 
uncertainties remain.  Any accounting of the present-day distribution of 
baryons must include an accurate census of these clouds and the mass 
associated with them.  As discussed in greater detail below (\S3),
precision measurements of the amount of warm (photoionized) IGM will
take considerable effort.  Not only must we characterize the absorber
distribution in H~I column density, but we must also apply a large 
ionization correction for the amount of unseen hydrogen in 
ionized form.  This correction is straightforward, depending 
on the ratio, $J_0/n_H$, of ionizing background intensity to gas
density.  Although the gas density, $n_H$, is loosely related to the 
column density, N$_{\rm HI}$, it is more accurately derived from the absorber 
geometry, including the characteristic scale length for variations
in the gas distribution.   

We would also like to use the H~I line widths and Doppler parameters 
to understand the thermodynamical properties of the IGM.  What is
its temperature?  When was energy deposited into the IGM
from photoionization or bulk outflows?  The Ly$\beta$/Ly$\alpha$ 
curves of growth yield reliable H~I column densities of 
the saturated Ly$\alpha$ lines with N$_{\rm HI} > 10^{13.5}$ cm$^{-2}$.  
Present estimates of the distribution in N$_{\rm HI}$ together with the
expected photoionization correction (H$^+$/H$_{\rm tot}$) suggest that 
the baryon content is dominated by the high end of the H~I column-density 
distribution.  In addition, observations of Ly$\beta$ lines corresponding 
to the Ly$\alpha$ absorbers confirms the identification of any Ly$\alpha$ 
systems that may have been confused with metal lines intrinsic to the AGN.

\subsection{Results of the HST Survey} 

\begin{figure}[h]
\begin{center}
\epsfxsize=5in
\leavevmode
\epsffile{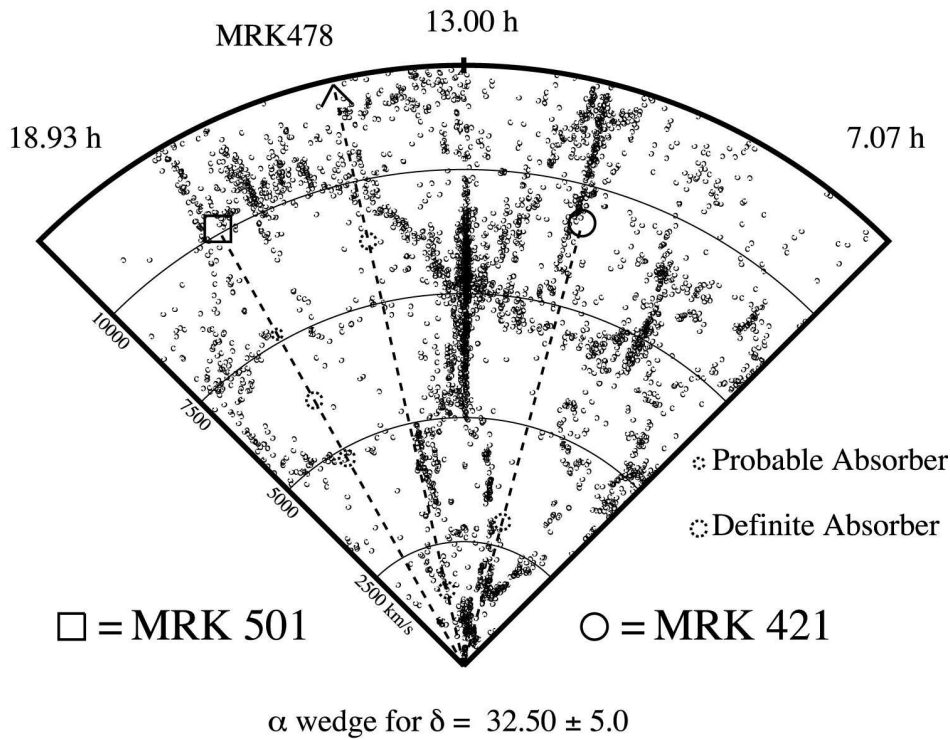}
\end{center}
\caption{\small
Locations of 3 sightlines used to study the low-$z$ IGM toward
Mrk~421, Mrk~501, and Mrk~478, in relation to large-scale filaments
and voids in the distribution of bright galaxies (CfA survey).
The positions of Ly$\alpha$ absorbers are shown as dotted circles,
along sightlines to background sources. }
\end{figure}

Our HST data were taken toward AGN (quasars, Seyferts, BL Lacs) 
brighter than $B = 15.5$ mag, corresponding to UV spectral flux 
$F_{\lambda} \geq (1-2) \times 10^{-14}$ ergs cm$^{-2}$ s$^{-1}$ \AA$^{-1}$.  
Most of our targets were chosen to lie behind well-studied galaxy 
distributions, in order to probe the connection of the Ly$\alpha$ absorbers 
with large-scale distributions of galaxies and voids (see Figure 2 and the
paper by Stocke in this volume). Our GHRS survey (Penton et al.\ 2000a,b) 
included 81 Ly$\alpha$ absorbers with line significance greater than
$4 \sigma$, along 15 sightlines.  With STIS, we have added 16 more 
AGN sightlines, bringing the survey total to 197 Ly$\alpha$ absorbers 
over a cumulative  pathlength 
$\Delta z \approx 1.15$. Our survey yields a line frequency,
$d{\cal N}/dz \approx 200$ for N$_{\rm HI} \geq 10^{13}$ cm$^{-2}$  
at $z < 0.1$. The mean distance between such absorbers along the sightline is 
\begin{equation}
     \langle \ell_{Ly\alpha} \rangle  = \frac {c/H_0}{d{\cal N}/dz } 
 \approx (20~{\rm Mpc})h_{75}^{-1}  \; ,
\end{equation}  
for a Hubble constant $H_0 = (75~{\rm km~s}^{-1}~{\rm Mpc}^{-1}) h_{75}$. 
This mean distance decreases by approximately a factor of 2, when we
include the more numerous weak Ly$\alpha$ absorbers (10 -- 50 m\AA) with 
column densities $12.3 \leq N_{\rm HI} < 13.0$. 

\begin{figure}[h]
\begin{center}
\epsfxsize=3.5in
\leavevmode
 \epsffile{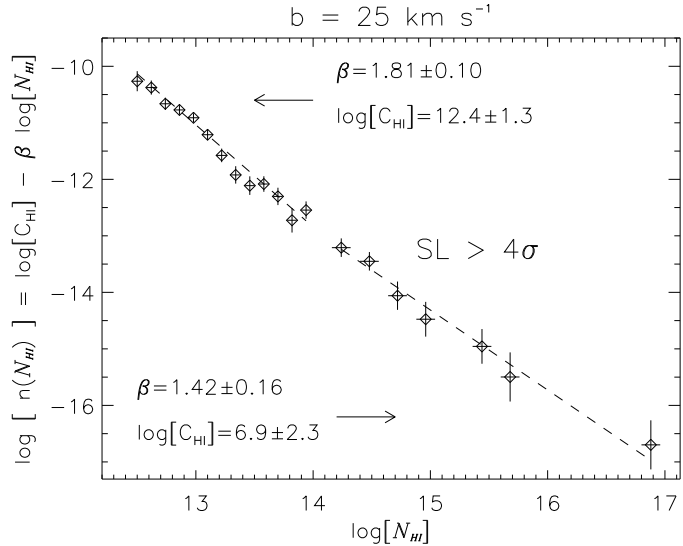}
\end{center}
 \caption{\small
   The distribution of H~I column densities in the combined GHRS/STIS
   survey (Penton et al.\ 2001b) for 197 Ly$\alpha$ absorbers at
   significance level $4\sigma$ or greater.  The weak absorbers
   fit a power law N$_{\rm HI}^{-1.81 \pm 0.10}$ while absorbers
   above $10^{14}$ cm$^{-2}$ follow a distribution 
   N$_{\rm HI}^{-1.42 \pm 0.16}$.  } 
\end{figure}

In our GHRS/STIS survey of 197 Ly$\alpha$ absorbers (Penton et al.\ 2001b),
the Ly$\alpha$ rest-frame equivalent widths range from 
$W_{\lambda} \approx 10$ m\AA\ to just above 1 \AA. 
For unsaturated Ly$\alpha$ lines, 
N$_{\rm HI} = (9.2 \times 10^{12}~{\rm cm}^{-2})(W_{\lambda}/50~{\rm m\AA})$.
The H~I column densities were derived using 
curves of growth (CoG) with Doppler parameters $b = 25 \pm 5$ km~s$^{-1}$.  
The $W_{\lambda}$ distribution exhibits a significant break at
$W_{\lambda} < 133$ m\AA, with an increasing number of weak absorbers
(10 -- 100 m\AA). These weak Ly$\alpha$ absorbers provide the most numerous 
and extensive probes of low-density regions of the IGM.
We characterize the distribution in H~I column densities (Figure 3) as 
a power law,
\begin{equation}
   d {\cal N}/d N_{\rm HI} \propto N_{\rm HI} ^{-\beta} \; .  
\end{equation}
At the low-column end, the slope is somewhat steeper, 
$\beta = 1.82 \pm 0.10$ ($12.3 < \log N_{\rm HI} < 14.0$),
while $\beta = 1.42 \pm 0.16$ for the saturated Ly$\alpha$ absorbers
($14.2 < \log N_{\rm HI} < 15.6$).  These values are similar to those 
found in the high-redshift Ly$\alpha$ forest (Kim et al.\ 1997, 2001).
Figure 4 shows the evolution of strong and weak Ly$\alpha$
absorbers with redshift, over the interval $0 < z < 3.7$.

\begin{figure}[h]
\begin{center}
\epsfxsize=3.5in
\leavevmode
\epsffile{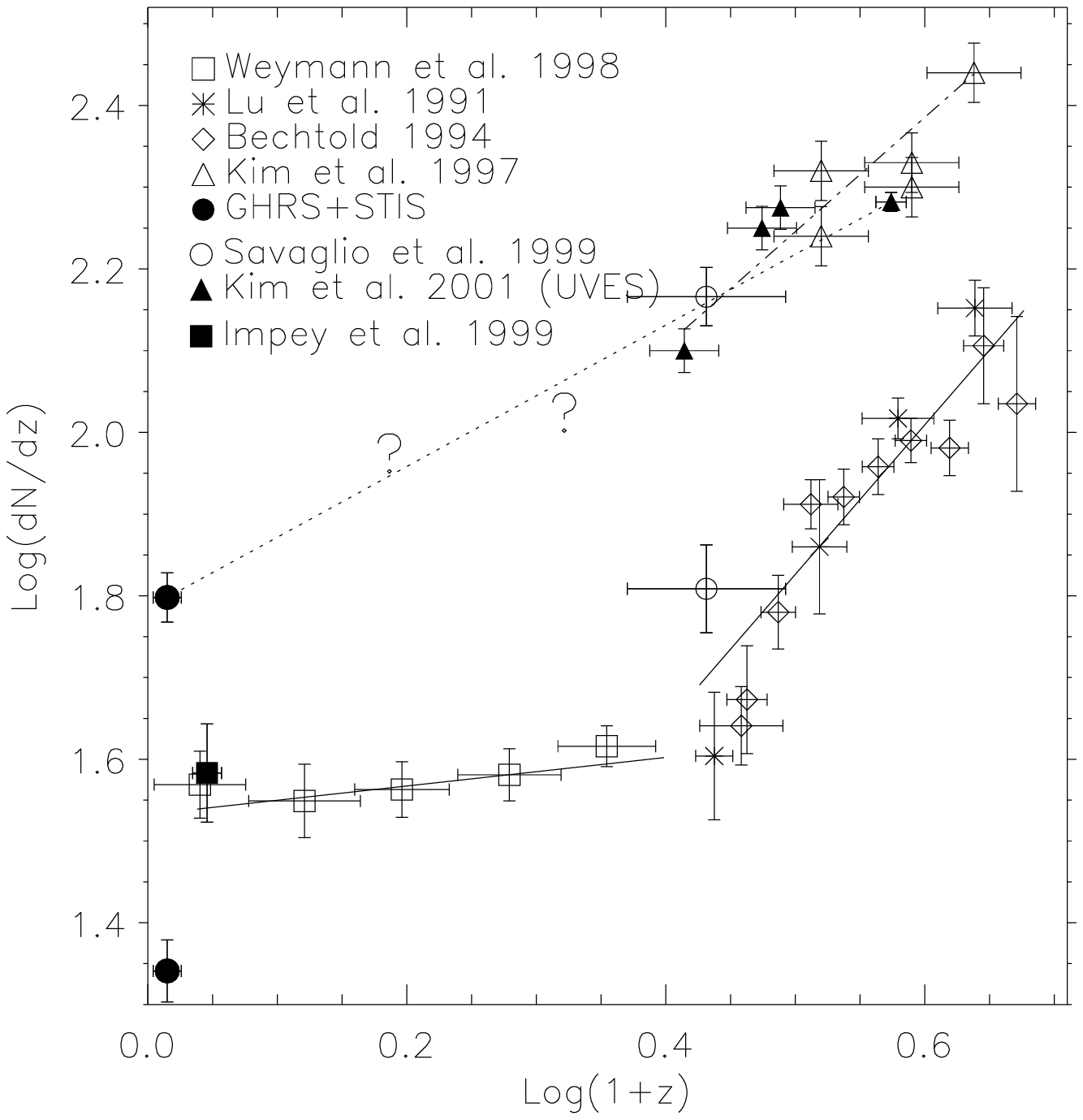}
\end{center}
\caption{\small
   Distribution of the line frequency, $d{\cal N}/dz$, of 
   strong and weak Ly$\alpha$ absorbers with redshift.  Results 
   from our GHRS/STIS survey are shown as solid filled circles at 
   $z \approx 0$ (Penton et al. 2001b). Top curve shows weak absorbers
   ($13.1 \leq$ log N$_{\rm HI} \leq 14.0$) from Keck/HIRES and VLT/UVES 
   (Kim et al.\ 1997, 2001) at $z > 1.6$, extrapolated down to the 
   $z \approx 0$ point (GHRS/STIS). Bottom curve shows
   the evolution of strong absorbers ($W_{\lambda} \geq 240$ m\AA\
   or log N$_{\rm HI} \geq 14$)  
   from ground-based surveys (Bechtold et al.\ 1994; Savaglio et al.\
   1999), extrapolated below $z < 1.5$ with the HST
   Key Project (Weymann et al.\ 1998), a low-resolution HST/GHRS/G140L 
   study (Impey, Petry, \& Flint 1999), and down to $z \approx 0$ with 
   our moderate-resolution GHRS/STIS survey. }
\end{figure}

\subsection{FUSE Studies of the Ly$\beta$ Forest}

\begin{figure}[h]
\begin{center}
\epsfxsize=5in
\leavevmode
\epsffile{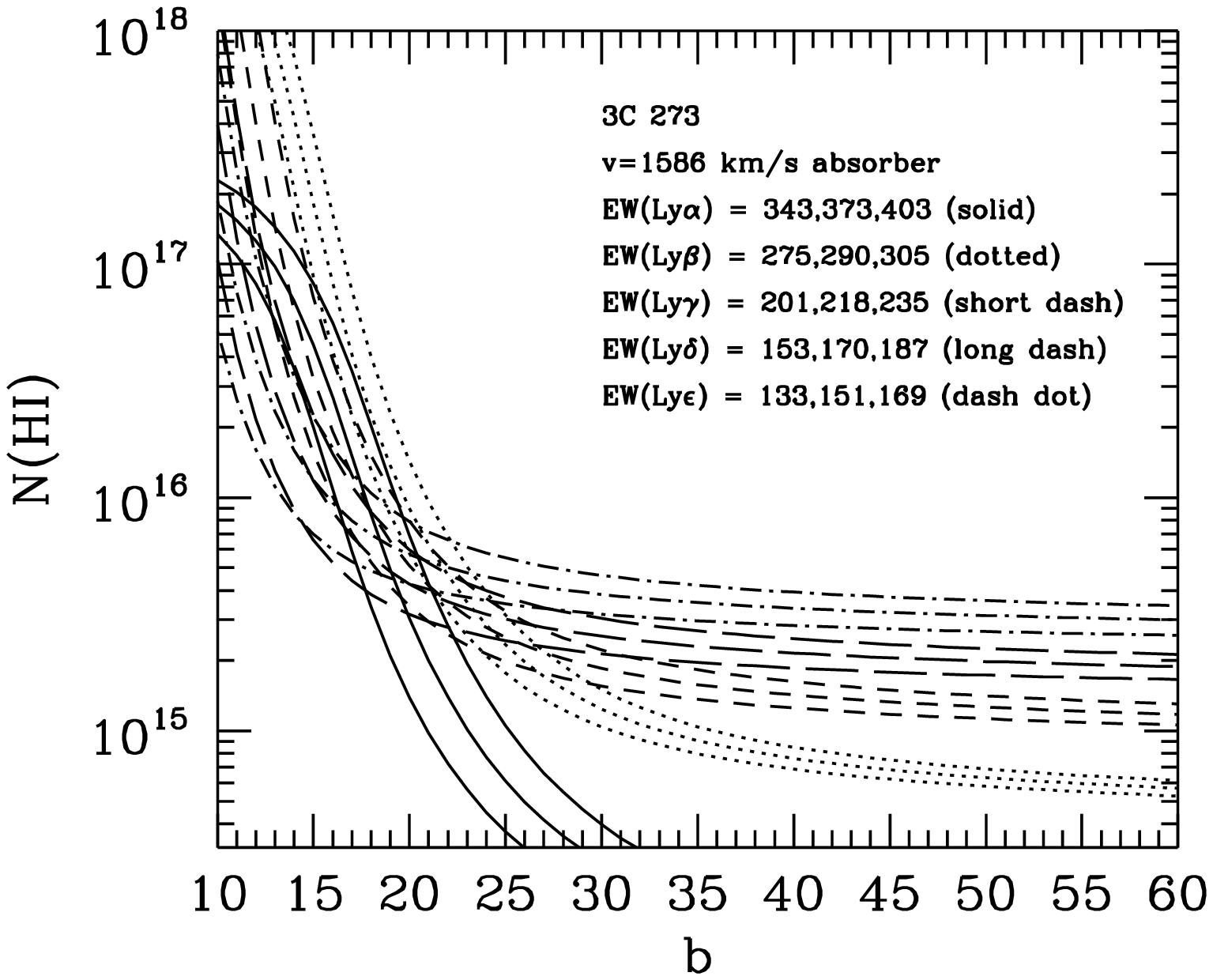}
\end{center}
\caption{\small
Curve of growth concordance plot (Giroux \& Shull, unpublished)
in $b$ (km~s$^{-1}$) and N$_{\rm HI}$ (cm$^{-2}$) for the Ly$\alpha$, 
Ly$\beta$, and higher Lyman lines in the strong absorber at $cz = 1586$ 
km s$^{-1}$ toward 3C~273.
Three curves for each Lyman line show single-component, Doppler-broadened
CoG fits to equivalent widths (labeled in m\AA\ with $\pm 1 \sigma$ errors.)
The best fit for Ly$\beta$ and higher lines yields 
log~N$_{\rm HI} = 15.85^{+0.10}_{-0.08}$ and $b = 16.1 \pm 1.1$
km~s$^{-1}$ (Sembach et al.\ 2001), considerably different
from the values, log N$_{\rm HI} = 14.22 \pm 0.07$ and
$b = 34.2 \pm 3.3$ km~s$^{-1}$, from Ly$\alpha$ profile
fitting (Weymann et al.\ 1995). }
\end{figure}

One of the Early Release Observations of the FUSE program was a
moderate-resolution (20-25 km~s$^{-1}$) study of the low-redshift IGM
(Shull et al.\ 2000).  Our team carried out studies of 7
extragalactic sightlines and 12 Ly$\beta$ absorbers that correspond
to Ly$\alpha$ lines detected by HST/GHRS and STIS.  
In general, we detect Ly$\beta$ absorption for all Ly$\alpha$
systems with $W_{\lambda} > 200$ m\AA.   This is not surprising,
since for unsaturated lines, the equivalent widths are:
$W_{\lambda}({\rm Ly}\alpha) = (54.5~{\rm m\AA}) 
(N_{\rm HI}/10^{13}~{\rm cm}^{-2})$ and $W_{\lambda}({\rm Ly}\beta) 
= (7.37~{\rm m\AA}) (N_{\rm HI}/10^{13}~{\rm cm}^{-2})$.
Even considering line saturation, the Ly$\beta$ line should
have equivalent width $W_{\lambda}({\rm Ly}\beta) \geq 
W_{\lambda}({\rm Ly}\alpha)/7.4$.    
Using FUSE data, with 30--40 m\AA\ (4$\sigma$) Ly$\beta$ detection limits, we 
employed the equivalent width ratio of Ly$\beta$/Ly$\alpha$ and occasional 
higher Lyman lines to determine the Doppler parameters, $b_{\rm CoG}$, 
and column densities, N$_{\rm HI}$, for moderately saturated lines. 
The Ly$\beta$/Ly$\alpha$ technique is demonstrated in Figure 5,
which shows the CoG concordance for the strong 1586 km~s$^{-1}$ 
Ly$\alpha$ absorber toward 3C~273.  In this case 
(Sembach et al.\ 2001), we detect the 
first 8 Lyman lines (Ly$\alpha$ through Ly$\theta$).  Surprisingly,
the CoG-inferred column density and $b$-value differ considerably 
from those derived from Ly$\alpha$ profile fitting.   

From a CoG analysis, the Ly$\beta$/Ly$\alpha$ ratios in our FUSE survey 
(Shull et al.\ 2000) yield a preliminary distribution function of Doppler 
parameters, with mean $\langle b \rangle = 31.4 \pm 7.4$ km~s$^{-1}$ and 
median $b = 28$ km~s$^{-1}$, comparable to values at redshifts $z = 2-3$. 
If thermal, these $b$-values correspond to $T_{\rm HI} = (m_H b^2/2k) 
\approx 50,000$~K, too hot for purely photoionized clouds
(Donahue \& Shull 1991). However, we find some evidence that the line widths
are not entirely thermal.  The CoG-inferred Doppler parameters are 
considerably less than the widths derived from Ly$\alpha$ 
profile fitting, $\langle b_{\rm CoG}/ b_{\rm width} \rangle  = 0.52$. 
The combined HST/FUSE data suggest that the low-$z$ Ly$\alpha$ absorbers 
contain significant non-thermal motions or velocity components in the 
line profile, perhaps arising from cosmological expansion and infall.

Because the CoG generally produces lower $b$-values, the derived
H~I column densities increase.  The typical increase over that derived from 
Ly$\alpha$ profile fitting is $\Delta[\log {\rm N}_{\rm HI}] = 0.3$, 
but it can increase by more than a factor of 10. In an extreme case
(Figure 5) the 1586 km~s$^{-1}$ absorber toward 3C~273 increased in column
density by a factor of over 40, to log N$_{\rm HI} = 15.85^{+0.10}_{-0.08}$
(Sembach et al. 2001) compared to the value, log~N$_{\rm HI} = 14.22 
\pm 0.07$, determined from Ly$\alpha$ profile fitting (Weymann et al.\ 
1995).  This large change in N$_{\rm HI}$ arose because the
curve of growth gave a Doppler parameter $b_{\rm CoG} = 16$ km~s$^{-1}$,
while Ly$\alpha$ profile fitting gave $b_{\rm width} = 34.2 \pm 3.3$
km~s$^{-1}$.

\subsection{Metallicities}

Determining the intergalactic heavy-element abundance (metallicity) of 
ionization state ($i$) of element ($Z$) depends in the first instance
on measuring accurate column densities, N$_{\rm HI}$ and N$_{Z,i}$. As 
discussed above, N$_{\rm HI}$ can be particularly sensitive to CoG 
effects arising from line saturation; 
the combination of Ly$\alpha$ and higher Lyman lines is often 
needed to fix the $b$-value. One must then apply an ionization correction,
to convert the ratio N$_{Z,i}$/N$_{\rm HI}$ to a total abundance
N$_{Z}$/N$_{H}$. If the absorbing gas is photoionized,
this ionization correction is straightforward (Donahue \& Shull
1991;  Shull et al.\ 1998) if one knows the intensity and spectrum
of the metagalactic background radiation and the physical density, $n_H$, 
of the absorber.  In practice, one computes the ionization correction
as a function of the dimensionless ``photoionization parameter", 
$U = n_{\gamma}/n_H$, where $n_{\gamma}$  is the number density 
(cm$^{-3}$) of Lyman-continuum photons and $n_H$ is the number density 
of hydrogen nuclei.  The situation becomes more complicated if 
the absorber's physical conditions include collisional ionization
in hotter gas, such as the O~VI absorbers seen by Tripp, Savage, \& 
Jenkins (2000).     

Metallicity limits in several low-$z$ Ly$\alpha$ clouds have  
been set for the strong 17,000 km~s$^{-1}$ absorbers toward 
PKS~2155-304 (Shull et al.\ 1998). The absence of detectable 
Si~III $\lambda 1206.5$ or C~IV $\lambda 1548.2$ sets limits of less 
than 0.003 solar. Other promising systems Ly$\alpha$ systems have
been found, including two strong Ly$\alpha$ absorbers 
toward PG~1211+143 (see Figure 6). With rest-frame equivalent widths 
of 1.14 \AA\ and 0.89 \AA, these absorbers are among the strongest
detected in our GHRS/STIS sample, and they appear to show heavy
elements at the level of several percent solar metallicity.     
Additional studies of Ly$\alpha$ absorbers with $W_{\lambda} > 200$ m\AA\
are currently underway with both FUSE and HST. 

\begin{figure}[h]
\begin{center}
\epsfxsize=5in
\leavevmode
\epsffile{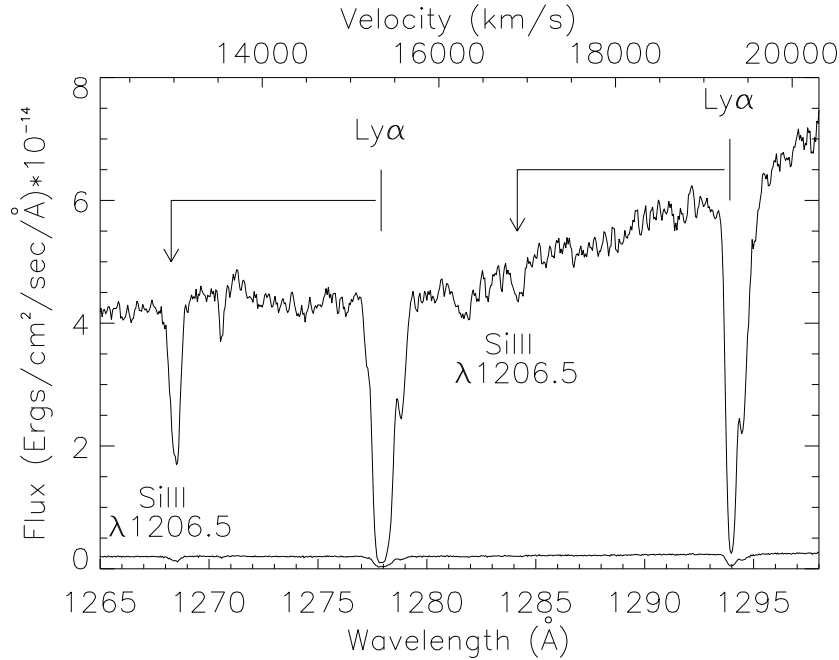}
\end{center}
\caption{\small
Section of our STIS/G140M observations of PG~1211+143, showing
two strong Ly$\alpha$ absorbers at $cz = 15,300$ and 19,400 km~s$^{-1}$,
with rest equivalent widths of $1145 \pm 115$ m\AA\ and $893 \pm 90$ m\AA,
respectively. Each of these absorbers appears to have multiple
velocity components.  The two arrows show the expected locations of the
Si~III $\lambda1206.50$ line, although the feature at 1268.3 \AA\ 
is actually another Ly$\alpha$ absorber (confirmed by Ly$\beta$).  } 
\end{figure}

\section{Toward Precision Measurements of the IGM Baryon Content}  

The primary ingredients for making accurate measurements of
the contribution of low-$z$ Ly$\alpha$ absorbers to $\Omega_b$ are: 
\begin{itemize}

\item Accurate H~I column densities, derived from a CoG analysis of 
   Ly$\alpha$, Ly$\beta$, and occasional higher Lyman lines. 

\item Knowledge of the ionizing radiation field, neeeded to derive the
    photoionization corrections for total abundances. 

\item An estimate of the physical density, $n_H$, from independent
    determinations of H~I absorber sizes, shapes, and gas distribution. 

\end{itemize} 
Toward these ends, we are attempting to accumulate a sufficiently
large database of Ly$\alpha$ and Ly$\beta$ absorbers with HST and FUSE
to characterize the distribution in N$_{\rm HI}$.  However well one can 
determine the N$_{\rm HI}$  distribution, one still must deal with
absorber geometric issues that produce systematic uncertainties 
in the ionization corrections. 

One of the best ways to make progress is to find a ``constellation
of AGN" whose UV spectra can be cross-correlated for common Ly$\alpha$ 
absorption lines to infer their characteristic sizes.  
At the current flux limits of the
2.4m HST, this is a difficult experiment, but there is
hope with the planned (Jan. 2004) installation of the 
{\it Cosmic Origins Spectrograph} (COS) on HST.  With 10--20 times 
greater throughput than STIS, COS will be able to obtain moderate
resolution (15--20 km~s$^{-1}$) spectra toward AGN as faint as 18 mag.  
As shown in Table 1, this greatly increases the chances of finding 
multiple bright QSOs separated by $\sim 30'$.  With a sufficient
number of AGN background sources behind low-$z$ Ly$\alpha$
absorbers, one can make rudimentary ``tomographic maps" of the 
cosmic web of warm (photoionized) baryons
left over from the epoch of large-scale structure formation.  

Mapping the evolution of these gaseous structures down to low $z$ 
is a prime scientific goal of COS. The full experiment must 
await the powerful 6--8m successor to HST, a mission concept known as 
the {\it Space Ultraviolet-Visible Observatory} (SUVO).  The scientific
and technological rationale for SUVO is contained in the ``White
Paper" from the UV-Optical Working Group (Shull et al.\ 1999c)
available on the Web at http://origins.colorado.edu/uvconf/UVOWG.html.

\begin{table}[h]
  \begin{center}
{\bf Table 1} \\
{\bf QSO Counts and Mean Angular Distance$^{a}$ Between QSOs}  \\
\ \\
  \begin{tabular}{ccc}
  \hline \hline
  $m_B$      &   $N_{\rm QSO}$   &  $\theta_{\rm QSO}$  \\
  (magn)     &   (sqdeg$^{-1}$)  &   (arcmin)           \\
\hline
    &        &           \\
16  &  0.01  &  $300'$   \\
17  &  0.13  &  $ 83'$   \\
18  &  1.1   &  $ 29'$   \\
19  &  5.3   &  $ 13'$   \\
20  &  17    &  $7.3'$   \\
21  &  41    &  $4.7'$    \\
\hline
\end{tabular}
\end{center}
$^{a}$ Mean angular distance $\theta_{\rm QSO} = (1/2) N_{\rm QSO}^{-1/2}
    = (30')N_{\rm QSO}^{-1/2}$ between Poisson-distributed QSOs of
    frequency $N_{\rm QSO}$ per square degree (see Shull et al.\ 1999c).    

\end{table}

To thoroughly map the cosmic web, 
we need to observe QSOs at magnitudes down to $m_B \approx 20-21$, where
the the mean angular distance between QSOs on the sky is $\leq 10'$,
allowing for lower UV continuum fluxes owing to potential extinction.
After accounting for ultraviolet absorption from Lyman-limit
systems,  Picard \& Jakobsen (1993) found a steep rate of
increase, $d(\log N)/d(\log F_{\lambda})
= 2.7 \pm 0.1$ for quasars in the flux range $10^{-14}$ down to
$10^{-16}$ ergs cm$^{-2}$ s$^{-1}$ \AA$^{-1}$ (approximately
$m_B = 15$ down to $m_B = 20$).  The current limit
of HST/STIS for moderate-resolution spectroscopy is
$m_B \approx 15.5$, while  HST/COS will take this limit
to $m_B \approx 18$.  Another order-of-magnitude improvement
is required to capitalize on the large increase in QSO populations
at magnitudes $m_B = 18 - 20$.

In the next several years, the GALEX mission is expected to identify 
large numbers of QSOs in the magnitude range $18 < m_B < 20$.  The 
Sloan survey will provide redshifts for many of these targets.  The 
task of mapping the IGM structures from $z = 2$ down to $z = 0$ will 
be a highlight of the SUVO program, if its spectrographs are 
designed with sufficient throughput to undertake a
major survey of sightlines at high spatial frequency.
The goal is to make an IGM baryonic survey on sub-degree
angular scales, comparable to that of the
MAP explorer and to the structure seen in galaxy surveys.
In doing so, we will connect the high-redshift seeds of galaxies and
clusters with the distributions of galaxies and IGM in the modern epoch,
at redshifts $z < 1$.

\acknowledgments

This work was supported at the University of Colorado by grants 
from the Space Telescope Science Institute, the FUSE Science Program, 
and the NASA Theoretical Astrophysics program.  I thank my Colorado 
colleagues, John Stocke, Steve Penton, Mark Giroux, Jason Tumlinson,
Massimo Ricotti, and Nick Gnedin for their contributions toward this 
research.

\end{document}